\documentclass[aps,showpacs,prx,superscriptaddress,preprint]{revtex4-1}%
\usepackage{amsfonts}
\usepackage{amsmath}
\usepackage{amssymb}
\usepackage{graphicx}
\usepackage{txfonts}
\usepackage{xcolor}%

\begin{document}

\title{Berry phases of higher spins due to internal geometry of Majorana constellation and relation to quantum entanglement}
\author{Chon-Fai Kam}
\affiliation{Department of Physics and The Hong Kong Institute of Quantum Information Science and Technology, 
The Chinese University of Hong Kong, Shatin, New Territories, Hong Kong, China}
\affiliation{Department of Mathematics, University of Macau, Avenida da Universidade, Taipa, Macau, China}
\author{Ren-Bao Liu}
\email{Email: rbliu@cuhk.edu.hk}
\affiliation{Department of Physics and The Hong Kong Institute of Quantum Information Science and Technology, 
The Chinese University of Hong Kong, Shatin, New Territories, Hong Kong, China}

\begin{abstract}
Majorana stars, the antipodal directions associated with the coherent states that are orthogonal to a spin state, provide a visualization and a geometric understanding of the structures of general quantum states. For example, the Berry phase of a spin-1/2 is given by half the solid angle enclosed by the close path of its Majorana star. It is conceivable that the Berry phase of higher spins may also be related to the geometry of the Majorana constellation. We find that for a spin-1 state, besides the expected contributions from the solid angles enclosed by the close paths of the two Majorana stars, the Berry phase includes a term related to the twist of the relative position vector around the barycenter vector of the two Majorana stars, i.e., the self-rotation of the constellation. Interestingly, if the spin-1 state is taken as a symmetrized two-qubit state, the extra contribution to the Berry phase is given by the self-rotation of the Majorana constellation weighted by the quantum entanglement of the two qubits. This discovery alludes to the relevance of the Majorana stellar geometry in representing the deep structures of quantum states and of quantum entanglement.
\end{abstract}

\maketitle
\textit{Introduction}. The Majorana constellation~\cite{majorana1932atomi} is a generalization of the Bloch representation in which a point on a unit sphere represents a spin-$1/2$ state. 
The Majorana stars are the $2j$ antipodal directions associated with the spin coherent states~\cite{perelomov2012generalized} that are orthogonal to a spin-$j$ state. 
The Majorana constellation provides an intriguing and useful visualization of general quantum states~\cite{bloch1945atoms, salwen1955resonance, meckler1958majorana, schwinger1977majorana, penrose1984spinors, penrose1960spinor, penrose1989emperor, zimba1993bell, hannay1996chaotic, hannay1998berry, hannay1998majorana, dennis2001topological, dennis2004canonical}. For example, the Majorana representation may be employed to analyze the collective modes and topological excitations of bosonic spinor atoms~\cite{barnett2006classifying, barnett2007classifying, makela2007inert, barnett2009geometrical, lian2012searching, cui2013synthetic}. Besides, in quantum metrology~\cite{giovannetti2011advances}, the most sensitive spin states under generic collective SU$(2)$ rotations are those that have the Majorna constellations in the form of platonic solids~\cite{kolenderski2008optimal, bjork2015extremal, bjork2015stars, bouchard2017quantum, chryssomalakos2017optimal, goldberg2018quantum}.

As an example for visualizing deep structures of quantum states, the Majorana representation was applied to understand the multipartite entanglement~\cite{martin2010multiqubit, ribeiro2011entanglement, wang2012nonlocality, bohnet2016quantumness, meill2017symmetric, neven2018entanglement}. A classification scheme of entanglement in symmetric states under stochastic local operations and classical communication was developed using the degeneracy pattern of Majorana stars~\cite{bastin2009operational, mathonet2010entanglement}. The maximally entangled symmetric state in terms of the geometric measure was identified by using the Majorana representation~\cite{aulbach2010maximally, markham2011entanglement}. An entanglement measure for symmetric states was designed using the barycenter of Majorana stars~\cite{ganczarek2012barycentric}. Multipartite symmetric states with maximal entropy of entanglement were identified by using the symmetry of the Majorana stars~\cite{giraud2015tensor, baguette2015anticoherence}. The genuine tripartite entanglement of non-symmetric three-qubit states was found represented by three distinct Majorana stars~\cite{kam2020three}.

Particularly relevant to this study, the Majorana representation has been applied to the study of quantum geometric phases. The Pancharatnam-Berry phase between three quantum states in a high-dimensional Hilbert space can be represented by the spherical triangles connecting the Majorana stars~\cite{tamate2011bloch, ogawa2015observation}. The quantum geometric phase can be mapped to a many-body Aharonov-Bohm phase by using the Majorana representation~\cite{bruno2012quantum}, and was shown to be represented by the sum of solid angles subtended by the Majorana star trajectories and the correlations between the stars~\cite{liu2014representation}.

The Berry phase of a spin-1/2 equals half the solid angle subtended by its trajectory on the Bloch sphere. Such a simple geometric formulation can be generalized to higher spins using the Majorana stars in some simple cases. For example, for a spin-$j$ state $|j,m\rangle$ with $m$ being the magnetic quantum number along a moving quantization direction, the Majorana stars are distributed at two antipodal points with $j\pm m$ stars opposite or parallel to the quantization direction, and the Berry phase is half the sum of solid angles subtended by all the Majorana star trajectories. For general spin-$j$ states, however, the Berry phases do not have such a simple geometric interpretation. It is natural to ask: what are the geometric meaning of the Berry phases of general higher spin states?

In this paper, we provide a partial, nonetheless intriguing, answer to this question. We show that the Berry phases of general spin-$j$ states, in addition to half the sum of the solid angles enclosed by the Majorana star trajectories, have terms related to the internal geometry of the Majorana constellation. Furthermore, when the spin-$j$ state is regarded as a symmetrized multi-qubit state, such internal geometry is closely related to the entanglement of the qubits. In particular, for a spin-$1$ state, the extra contribution to the Berry phase is given by the self-rotation angle of the Majorana constellation, weighted by the entanglement of the symmetrized states of two qubits.    

\textit{Berry phases of a spin-$j$ in the Majorana representation}.
A spin-$j$ state can be written as a symmetric tensor product of $N=2j$ spin-1/2 states~\cite{hannay1998berry}
\begin{equation}
|\psi\rangle=\frac{1}{\sqrt{N!A_N}}\sum_{\sigma}|{{\mathbf n}}_{\sigma(1)}\rangle\otimes\cdots\otimes|{{\mathbf n}}_{\sigma(N)}\rangle,
\label{eq_symmetrized}
\end{equation}
where $A_N\equiv \sum_{\sigma}\prod_k\langle {{\mathbf n}}_k|{{\mathbf n}}_{\sigma(k)}\rangle$ is a normalization factor, the summation is over all permutations $\{\sigma\}$, and $|{{\mathbf n}}_k\rangle$ is a spin-1/2 state polarized along the direction ${{\mathbf n}}_k$. The antipodal directions $-{{\mathbf n}}_k$ of the Majorana stars ${{\mathbf n}}_k$ correspond to the spin coherent states $|-{{\mathbf n}}_k,j\rangle\equiv |-{{\mathbf n}}_k\rangle^{\otimes N}$ that are orthogonal to the spin-$j$ state $|\psi\rangle$. 

\begin{figure}[tbp]
\includegraphics[width=0.9\columnwidth]{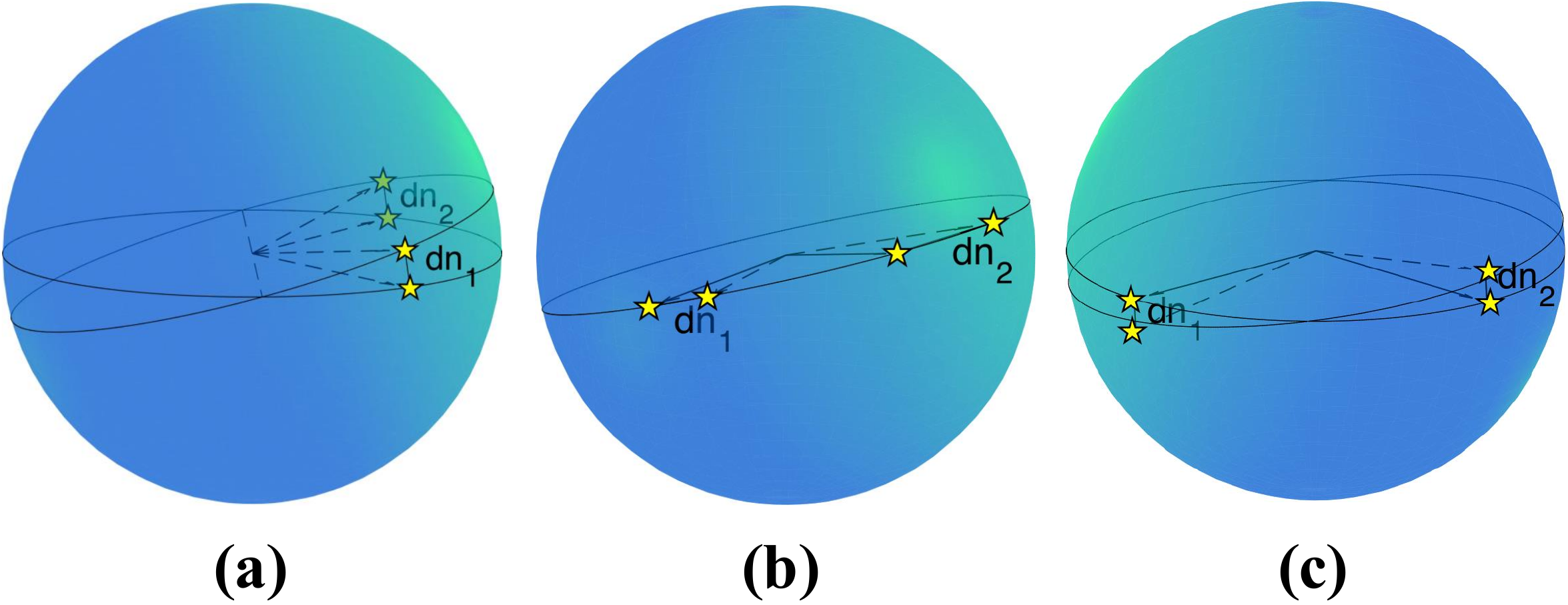}
\caption{Three cases for which the Berry phase is the sum of contributions from the Majorana star trajectories or contains an extra term. 
(a) A rigid rotation of the great circle that connects ${{\mathbf n}}_1$ and ${{\mathbf n}}_2$ without self-rotation ($d{{\mathbf n}}_1=d{{\mathbf n}}_2$, and perpendicular to ${{\mathbf n}}_1$ and ${{\mathbf n}}_2$).
 (b) ${{\mathbf n}}_1$ and ${{\mathbf n}}_2$ slide along the great circle. (c) Self-rotation of the constellation ($d{{\mathbf n}}_1=-d{{\mathbf n}}_2$, and perpendicular to ${{\mathbf n}}_1$ and ${{\mathbf n}}_2$).}
\label{MajoranaStarPlot}
\end{figure}

Let us begin with the simplest non-trivial case, {i.e.}, a spin-$1$ state with two Majorana stars, which may be written as the symmetric tensor product of two spin-1/2 states as $$|\psi\rangle = \frac{1}{\sqrt{2A_2}}\left(|{{\mathbf n}}_1\rangle\otimes|{{\mathbf n}}_2\rangle+|{{\mathbf n}}_2\rangle\otimes|{{\mathbf n}}_1\rangle\right),$$ with $A_2=1+|\langle {{\mathbf n}}_1|{{\mathbf n}}_2\rangle|^2$. A direct computation yields
\begin{align}\label{BerryConnection}
    \langle\psi|\dot{\psi}\rangle&=A_2^{-1}\left\{\langle {{\mathbf n}}_1|\dot{{{\mathbf n}}}_1\rangle+\langle {{\mathbf n}}_2|\dot{{{\mathbf n}}}_2\rangle+\frac{1}{2}\left(\langle {{\mathbf n}}_1|\dot{{{\mathbf n}}}_2\rangle-\langle \dot{{{\mathbf n}}}_1|{{\mathbf n}}_2\rangle\right)\langle {{\mathbf n}}_2|{{\mathbf n}}_1\rangle\right.\nonumber\\
    &\left.+\frac{1}{2}\left(\langle {{\mathbf n}}_2|\dot{{{\mathbf n}}}_1\rangle-\langle \dot{{{\mathbf n}}}_2|{{\mathbf n}}_1\rangle\right)\langle {{\mathbf n}}_1|{{\mathbf n}}_2\rangle
    \right\}.
\end{align}
The Berry connection $\mathcal{A}\equiv i\langle\psi|d\psi\rangle$~\cite{hannay1998berry} is
\begin{align}\label{Connection1Form}
   \mathcal{A}&=\sum_i\mathcal{A}({{\mathbf n}}_i)-\frac{1}{2}\frac{({{\mathbf n}}_1\wedge {{\mathbf n}}_2)\cdot(d {{\mathbf n}}_1-d {{\mathbf n}}_2)}{3+ {{\mathbf n}}_1\cdot  {{\mathbf n}}_2},
\end{align}
where $\mathcal{A}({{\mathbf n}}_i)=-\frac{1}{2}(1-\cos\theta_i)d\phi_i$ is the Berry connection of the $i$-th Majorana star.

The second term in Eq.\:\eqref{Connection1Form} is related to the internal geometry of the Majorana constellation. It vanishes in two cases: (i) $d{{\mathbf n}}_1-d{{\mathbf n}}_2=0$; (ii) $d({{\mathbf n}}_1-{{\mathbf n}}_2)\perp{{\mathbf n}}_1\wedge {{\mathbf n}}_2$. For case (i), the great circle that connects the two Majorana stars ${{\mathbf n}}_1$ and ${{\mathbf n}}_2$ undergoes a rigid rotation without self-rotation of the constellation (see Fig.~\ref{MajoranaStarPlot}a); For case (ii), the two Majorana stars slide along the great circle connecting them (see Fig.~\ref{MajoranaStarPlot}b). The case ${{\mathbf n}}_1=\pm{{\mathbf n}}_2$ is a special case of (ii). In these two cases, the Berry phases would be the half sum of the solid angles enclosed by the trajectories of the Majorana stars. But in general, the Berry phases would contain an extra term due to the last term in Eq.\:\eqref{Connection1Form} (see Fig.~\ref{MajoranaStarPlot}c for an example).

\begin{figure}[tbp]
\begin{center}
\includegraphics[width=0.5\columnwidth]{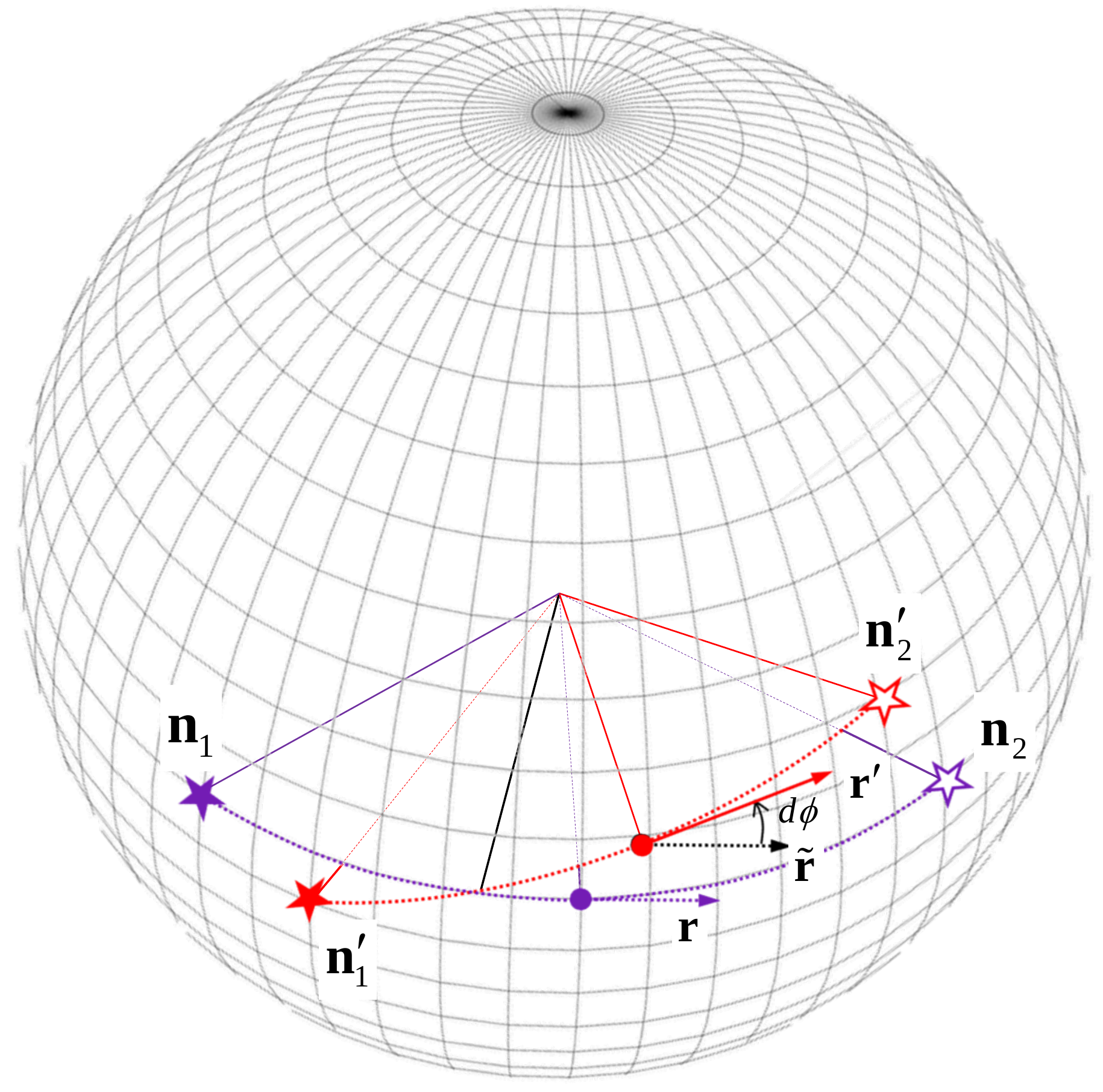}
\end{center}
\caption{Illustration of the differential self-rotation $d\phi$ of a constellation due to the motion from $({\mathbf n}_1,{\mathbf n}_2)$ to $({\mathbf n}'_1,{\mathbf n}'_2)$, where $\tilde{\mathbf r}$ is ${\mathbf r}$ parallel-transported from $\hat{\mathbf R}$ to $\hat{\mathbf R}'$. The angle from $\tilde{\mathbf r}$ to ${\mathbf r}'$ is the self-rotation angle.}
\label{fig:rotation}
\end{figure}

\textit{Self-rotation of the constellation}. We define the barycenter ${{\mathbf R}}\equiv \frac{1}{2}({{\mathbf n}}_1+{{\mathbf n}}_2)$ and the relative displacement ${{\mathbf r}}\equiv {{\mathbf n}}_1-{{\mathbf n}}_2$, and rewrite Eq.\:\eqref{Connection1Form} as
\begin{equation}\label{PrincipalFormula}
    \mathcal{A}= \sum_i\mathcal{A}({{\mathbf n}}_i)+\mathcal{C}(\Theta)\cos\Theta\hat{{{\mathbf R}}}\cdot(\hat{\mathbf r}\wedge d\hat{\mathbf r}),
\end{equation}
where $2\Theta\equiv\arccos({{\mathbf n}}_1\cdot{{\mathbf n}}_2)$ is the angle between the two Majorna stars, $\mathcal{C}(x)\equiv\sin^2x/(1+\cos^2x)$
is a continuous, strictly increasing function in $[0,\pi/2)$ with $\mathcal{C}(0)=0$ and $\mathcal{C}(\pi/2)=1$, and $\hat{\mathbf R}$ and $\hat{\mathbf r}$ are unit vectors along ${\mathbf R}$ and ${\mathbf r}$, respectively. 
The quantity 
\begin{align}
\hat{{{\mathbf R}}}\cdot(\hat{\mathbf r}\wedge d\hat{\mathbf r})\equiv d\phi
\end{align} in Eq.\:\eqref{PrincipalFormula} has a straightforward geometric meaning: it is the angle between $\tilde{\mathbf r}$ and the new displacement ${\mathbf r}'$, i.e., the self-rotation of the constellation, where $\tilde{\mathbf r}$ is the displacement ${\mathbf r}$ parallel transported from $\hat{\mathbf R}$ to the new barycenter direction $\hat{\mathbf R}'$. See Fig.~\ref{fig:rotation} for an illustration. The Berry phase for a spin-$1$ state in the Majorana representation becomes
\begin{equation}\label{SpinOneBerryPhase}
    \gamma=\sum_i\gamma_i+\oint\mathcal{C}(\Theta)\cos\Theta d\phi,
\end{equation}
where $\gamma_i$ is the Berry phases of the $i$-th star.

\begin{figure}[tbp]
\begin{center}
\includegraphics[width=0.5\columnwidth]{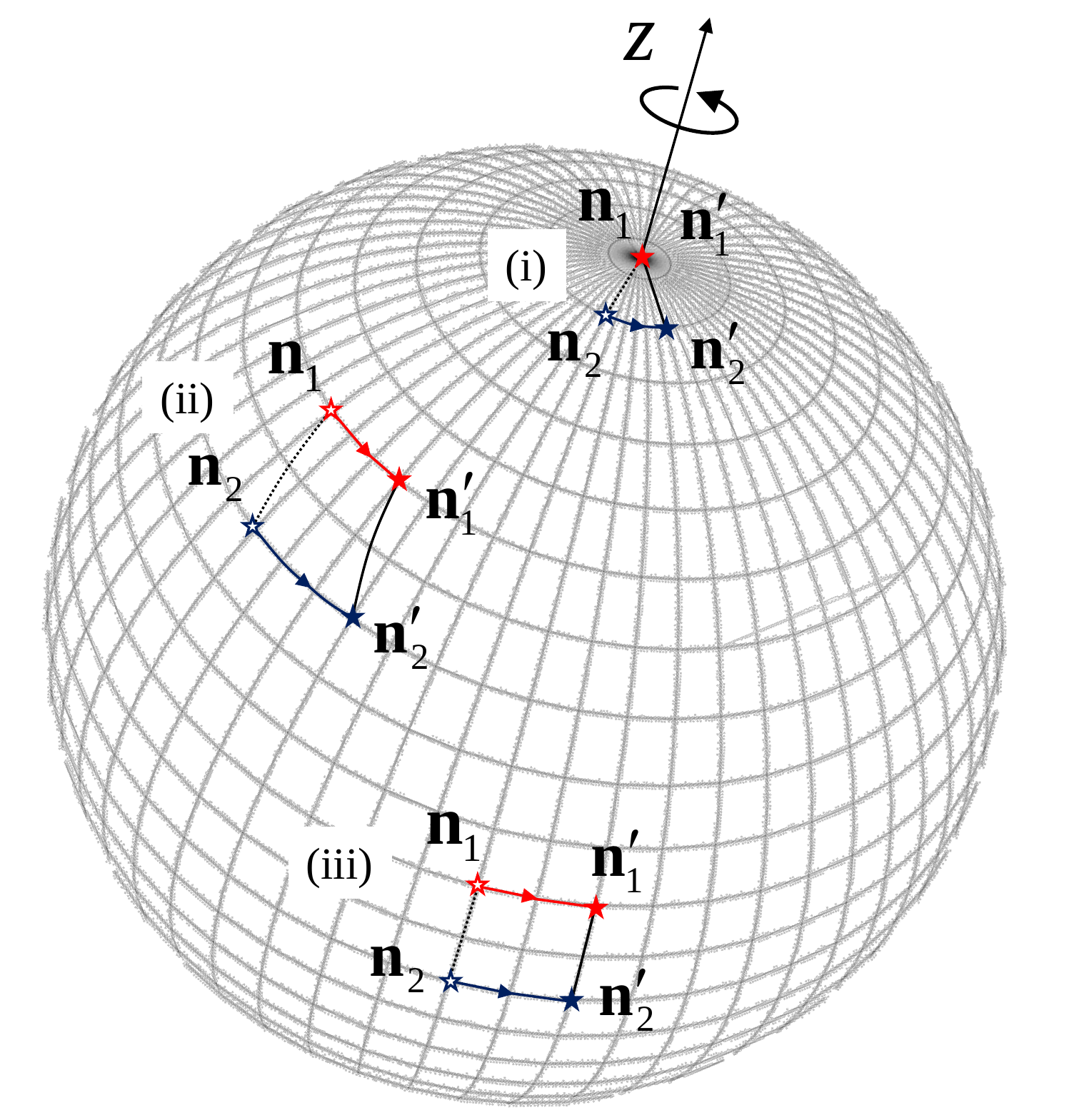}
\end{center}
\caption{Schematic of a pair of Majorana stars ${{\mathbf n}}_1$ and ${{\mathbf n}}_2$ located at different latitudes, rotating about the $z$-axis with the same angular frequency. (i) One star is at the north pole and another one is close to it. (ii) Two stars are away from the north pole and the equator. (iii) Two stars on the opposite sides of the equator at symmetric locations.}
\label{fig:example}
\end{figure}

\textit{Example}. We consider a simple example of two stars rotating about the $z$-axis with the same angular frequency (Fig.~\ref{fig:example}). The locations of the two stars can be simply written as ${\mathbf n}_{j}=\left(\sin\theta_{j}\cos(t+\phi_{j}),\sin\theta_{j}\sin(t+\phi_{j}), \cos\theta_{j}\right)$ with $j=1$ or $2$, $t$ varying from $0$ to $2\pi$, $\theta_{j}$ being the latitudes of the two stars, and $\phi_{j}$ being their initial longitudes. The latitude of the barycenter, i.e., the angle from the $z$ axis to $\hat{\mathbf R}$ is denoted as $\vartheta$. A direct calculation yields the differential self-rotation of the constellation
$$
d\phi=\hat{{{\mathbf R}}}\cdot(\hat{\mathbf r}\wedge d\hat{\mathbf r}) =\cos\vartheta dt,
$$
and the Berry phase in a period is
$$
\gamma=\pi\cos\theta_1+\pi\cos\theta_2+2\pi\cos\vartheta {\mathcal C}\cos\Theta.
$$
This result is consistent with the intuition --- when the constellation is located close to the north pole (case (i) in Fig.~\ref{fig:example}), the self-rotation angle is the same as the rotation angle of the constellation about the $z$-axis; when it is close to the equator (case (iii) in Fig.~\ref{fig:example}), the two stars move in parallel without self-rotation; and in general, the self-rotation is reduced by a factor $\cos\vartheta$ (case (ii) in Fig.~\ref{fig:example}).

\textit{General spin states}.
For a general spin-$j$ state, the Berry connection in the Majorana representation can be expressed as~\cite{liu2014representation}
\begin{equation}\label{General1Form}
\mathcal{A}=\sum_i\mathcal{A}({{\mathbf n}}_i)-\frac{1}{2A_N}\sum_{i<j}\frac{\partial{A_N}}{\partial{({{\mathbf n}}_i\cdot {{\mathbf n}}_j)}}( {{\mathbf n}}_i\wedge {{\mathbf n}}_j)\cdot(d {{\mathbf n}}_i-d {{\mathbf n}}_j).
\end{equation}
Here $A_N$ is an expansion of symmetric functions of ${{\mathbf n}}_i\cdot {{\mathbf n}}_j$ as~\cite{lee1988wehrl} $A_N = [(N+1)!/2^N]\sum_{k=0}^{[N/2]}[D_k^N/(2k+1)!!]$ with $D_k^N\equiv {\sum}_{i_1=1}^N{\sum}_{j_1>i_1}^N\cdots{\sum'}_{i_k>i_{k-1}}^{N}{\sum'}_{j_k>i_k}^{N}({{\mathbf n}}_{i_1}\cdot {{\mathbf n}}_{j_1})\cdots({{\mathbf n}}_{i_k}\cdot {{\mathbf n}}_{j_k})$, where $\sum'$ indicates that all non-repeated indices in each term take different values, and $D_0^N\equiv 1$. The Berry phase for a spin-$j$ state in the Majorana representation may be expressed as
\begin{equation}\label{GeometricBerryPhase}
    \gamma=\sum_i\gamma_i+\sum_{i<j}\oint w_{ij}d\phi_{ij},
\end{equation}
where $w_{ij}\equiv -\cos\Theta_{ij}\partial\ln{A_N}/\partial\ln{(2\sin^2\Theta_{ij})}$ is a weighting factor depending only on the angle $2\Theta_{ij}\equiv \arccos({{\mathbf n}}_i\cdot{{\mathbf n}}_j)$ between the $i$-th and $j$-th Majorana stars ${{\mathbf n}}_i$ and ${{\mathbf n}}_j$, and $d\phi_{ij}$ is the differential self-rotation of the two stars. When the Majorana constellation is rigid, {i.e.}, the states $|{{\mathbf n}}_i\rangle$ are subjected to collective SU$(2)$ rotations without changing their mutual distances, one immediately obtains $\gamma=\sum_i\gamma_i+\sum_{i<j} w_{ij}\phi_{ij}$. 

For spin-1 and higher spins, there is an interesting case that the Majorana stars can permute rather than completing their own cycles. As the Majorana stars are unordered points on the sphere, the final state of the system is the same as the initial one. The formulas Eqs.\:\eqref{SpinOneBerryPhase} and \eqref{GeometricBerryPhase} are unchanged, except that the loop integrals should be replaced by ordinary line integrals. For example, when $m$ and $2j-m$ coincident Majorana stars are in antipodal directions, and subjected to rigid rotations so as to remain antipodal, then the Berry phase is zero if the stars complete their own cycles, and is $(-1)^j$ if the stars exchange positions. This can be directly obtained from Eq.\:\eqref{GeometricBerryPhase} as the additional term in the Berry phase vanishes.

\textit{Relation to entanglement}. 
The spin-$j$ state can be regraded as a symmetrized multi-qubit state (see Eq.\:(\ref{eq_symmetrized})). Below we show that the weighting factors $w_{ij}$ associated with the self-rotation contributions to the Berry phase are related to the entanglement between the qubits. For symmetric two-qubit states with two Majorana stars, the function $\mathcal{C}(\Theta)$ in Eq.\:\eqref{SpinOneBerryPhase} is a measure of entanglement called the concurrence~\cite{bennett1996mixed, hill1997entanglement}
\begin{equation}
    \mathcal{C}\equiv\frac{1-{{\mathbf n}}_1\cdot {{\mathbf n}}_2}{3+{{\mathbf n}}_1\cdot {{\mathbf n}}_2}=\frac{\sin^2\Theta}{1+\cos^2\Theta},
\end{equation}
which is an entanglement monotone for bipartite states, equals to $0$ for separable states and $1$ for maximally entangled states. $\cos\Theta$ is related to another measure of entanglement, namely, the barycentric measure of entanglement $E_B$~\cite{ganczarek2012barycentric}, $\cos\Theta=\sqrt{1-E_B}$. Hence, the Berry phase for a spin-$1$ state in the Majorana representation can be expressed as
\begin{equation}\label{TwistedRibbonTopologies}
    \gamma=\sum_i\gamma_i+\oint\mathcal{C}\sqrt{1-E_B}d\phi.
\end{equation}
The additional term equals the integration of the self-rotation of the Majorana constellation, weighted by the concurrence and barycentric measures of the two-qubit entanglement. If the Majorana constellation undergoes rigid rotation, the entanglement is invariant and the Berry phase becomes
\begin{equation}\label{FinalRibbonTwist}
\gamma=\sum_i\gamma_i+ {\mathcal C}\sqrt{1-E_B}\phi.
\end{equation}

For symmetric three-qubit states with three distinct Majorana stars, $A_3=3+\sum_{i<j}{{\mathbf n}}_i\cdot{{\mathbf n}}_j$ and $w_{ij}=\sin^2\Theta_{ij}\cos\Theta_{ij}/\sum_{i<j}\cos^2\Theta_{ij}$. As the barycentric entanglement measure $E_B=\frac{4}{9}\sum_{i<j}\sin^2\Theta_{ij}$, and the three-tangle $\tau_3=\frac{4}{3}(\prod_{i<j}\sin\Theta_{ij}/\sum_{i<j}\cos^2\Theta_{ij})^2$, the product of the weighting factors is bounded by the entanglement between the qubits, $\prod_{i<j}w_{ij}\leq \frac{1}{4}\tau_3(1-\frac{3}{4}E_B)^{1/2}$. As a specific example, for states with Majorana stars distributed evenly on the Majorana sphere, {i.e.}, $\Theta_{12}=\Theta_{13}=\Theta_{23}\equiv\Theta$, one obtains $w_{12}=w_{23}=w_{13}=\frac{1}{3}\sin^2(\frac{\Theta}{2})/\cos(\frac{\Theta}{2})$, $E_B=\frac{4}{3}\sin^2\Theta$, and $\tau_3=\frac{4}{27}\sin^6\Theta/\cos^4\Theta$. Hence, when subjected to rigid rotation, the additional Berry phase can be written as $\Delta\gamma=\frac{1}{2}(\tau_3E_B)^{1/4}\sum_{i<j}\phi_{ij}$,  the sum of self-rotation angles of all Majorana star pairs weighted by the entanglement between the qubits.

As a remark, the additional term in the Berry phase is interpreted as an integration of pair solid angles weighted by entanglement in Ref.~\cite{liu2014representation}. However, the geometric meaning of the pair solid angles is vague, as it involves two different frames of reference. Here, our result provides a transparent geometric understanding of the additional term in the Berry phase.

\textit{Conclusion}.--- We analyzed the Berry phases for permutation symmetric $n$-qubit states using Majorana's star representation. We show that the Berry phase, in addition to the solid angles subtended by the trajectories of individual stars, contains an addition term related to the internal geometry of the Majorana constellation. In particular, for symmetric two-qubit states, the additional Berry phase is the integration of the self-rotation angle of the constellation weighted by the concurrence and barycentric entanglement measure. For higher spins, the internal geometry term can still be written as the sum of pairwise self-rotation with pairwise weighting factors, but, except for special cases, the explicit relations between the weighting factors and the multi-partite entanglement are still unclear. Conversely, it is interesting to examine whether the internal geometric properties of the Majorana stars may be employed to characterize the genuine multi-partite entanglement of permutation-symmetric states. This paper presents an interesting example of using Majorana constellations to characterize the structures of quantum states. We believe the Majorana representation is worth further exploration for understanding the deep structures of quantum states and Hilbert spaces.
 
\begin{acknowledgements}
This work was supported by Hong Kong RGC/GRF Project 14304117.
\end{acknowledgements}

\end{document}